\documentclass[10pt,prd,nofootinbib]{revtex4}
\usepackage{amsmath,amscd,amssymb}
\usepackage{color,epsfig,eso-pic}
\usepackage{xfrac}
\usepackage{latexsym}
\usepackage{mathrsfs}
\usepackage{graphicx}
\usepackage{bbm}      

\newcommand{\zr}[1]{\mbox{\hspace*{#1em}}}
\newcommand{\ID}{\mbox{{\sf 1}\zr{-0.16}\rule{0.04em}{1.55ex}\zr{0.1}}}
\newcommand{\fract}[2]{{\textstyle\frac{#1}{#2}}}
\newcommand{\imu}{{\rm i}}

\begin{document}
\baselineskip16pt

\title{Emerging Translational Variance: Vacuum Polarization Energy 
of the $\mathbf{\phi^6}$ Kink}

\author{H. Weigel}

\affiliation{
Institute for Theoretical Physics, Physics Department, 
Stellenbosch University, Matieland 7602, South Africa}

\begin{abstract}
We propose an efficient method to compute the vacuum polarization energy 
of static field configurations that do not allow a decomposition into 
symmetric and anti-symmetric channels in one space dimension. In particular 
we compute the vacuum polarization energy of the kink soliton in the $\phi^6$ 
model. We link the dependence of this energy on the position of the center of 
the soliton to the different masses of the quantum fluctuations at negative and 
positive spatial infinity.
\end{abstract}

\maketitle

\section{Motivation}

It is of general interest to compute quantum corrections to classical
field configurations like soliton solutions that are frequently interpreted
as particles. On top of the wish list we find the energies that predict
particle masses. The quantum correction to the energy can be quite
significant because the classical field acts as a background that 
strongly polarizes the spectrum of the quantum fluctuations about it.
For that reason the quantum correction to the classical energy is called
vacuum polarization energy (VPE). Here we will consider the leading,
{\it i.e.} one loop, contribution.

Field theories that have classical soliton solutions in various topological 
sectors deserve particular interest. Solitons from different sectors have unequal
winding numbers and the fluctuation spectrum changes significantly from one
sector to the other. For example, the number of zero modes is linked to the
number of (normalizable) zero modes that in turn arise from the symmetries
that are spontaneously broken by the soliton. Of course, the pattern of 
spontaneous symmetry breaking is subject to the topological structure. On the 
other hand, the winding number is typically identified with the particle 
number. The prime example is the Skyrme model\cite{Skyrme:1961vq,Skyrme:1988xj}
wherein the winding number determines the baryon 
number\cite{Witten:1979kh,Adkins:1983ya}. Many properties of baryons have
been studied in this soliton model and its generalization in the 
past\cite{Weigel:2008zz}. More recently configurations with very large
winding numbers have been investigated\cite{Feist:2012ps} and these solutions
were identified with nuclei. To obtain a sensible understanding of the predicted 
nuclear binding energies it is, of course, important to consider the VPE, in 
particular when it is expected to strongly depend in the particle number. So 
far this has not been attempted for the simple reason that the model is not 
renormalizable. A rough estimate\cite{Scholtz:1993jg}\footnote{See 
Ref.\cite{Meier:1996ng} for a general discussion of the Skyrmion's quantum 
corrections and further references on the topic.} in the context of the 
H--dibaryon\cite{Jaffe:1976yi,Balachandran:1983dj} suggests that the VPE strongly 
reduces the binding energy of multi--baryon states.

As already mentioned, one issue for the calculation of the VPE is renormalization.
Another important one is, as will be discussed below, that the VPE is (numerically) 
extracted from the scattering data for the quantum fluctuations about the classical 
configuration\cite{Graham:2009zz}. Though this so--called {\it spectral method}
allows for a direct implementation of standard renormalization conditions it has 
limitations as it requires sufficient symmetry for a partial wave decomposition. 
This may not be possible for configurations with an intricate topological 
structure associated with large winding numbers.

The $\phi^6$ model in $D=1+1$ dimensions has soliton solutions with different 
topological structures\cite{Lohe:1979mh,Lohe:1980js} and the fluctuations do not 
decouple into a parity channels. The approach employed here is also based on scattering 
data but advances the spectral method such that no parity decomposition is required. 
We will also see that it is significantly more effective than previous 
computations\cite{AlonsoIzquierdo:2002eb,AlonsoIzquierdo:2011dy,AlonsoIzquierdo:2012tw} 
for the VPE of solitons in $D=1+1$ dimensions that are based on heat kernel expansions 
combined with $\zeta$--function regularization 
techniques\cite{Elizalde:1996zk,Elizalde:1994gf,Kirsten:2000ad}. 

Although the $\phi^6$ model is not fully renormalizable, at one loop order the 
ultra--violet divergences can be removed unambiguously. However, another very interesting
phenomenon emerges. The distinct topological structures induce non--equivalent vacua 
that manifest themselves via different dispersion relations for the quantum fluctuations
at positive and negative spatial infinity. At some intermediate position the soliton
mediates between these vacua. Since this position cannot be uniquely determined the 
resulting VPE exhibits a translational variance. This is surprising since, after all, 
the model is defined through a local and translational invariant Lagrangian. In this paper 
we will describe the emergence of this variance and link it to the different level 
densities that arise from the dispersion relations. To open these results for 
discussion\footnote{The present paper reflects the author's invited presentation at the 
$5^{\rm th}$ {\it Winter Workshop on Non-Perturbative Quantum Field Theory} based on 
the methods derived in Ref.\cite{Weigel:2016zbs} making some overlap unavoidable.} it 
is necessary to review in detail the methods developed in Ref.\cite{Weigel:2016zbs} to 
compute the VPE for backgrounds in one space dimension that are not (manifestly) invariant 
under spatial reflection.

Following this introductory motivation we will describe the $\phi^6$ model and its kink 
solutions. In chapter III we will review the spectral method that ultimately
leads to a variant of the Krein--Friedel--Lloyd formula\cite{Faulkner:1977aa} for the VPE.
The novel approach to obtain the relevant scattering data will be discussed in chapter IV 
and combined with the one--loop renormalization in chapter V. A comparison with known 
(exact) results will be given in chapter VI while chapter VII contains the predicted VPE 
for the solitons of the $\phi^6$ model. Translational variance of the VPE that emerges 
from the existence of non--equivalent vacua will be analyzed in chapter VIII. We conclude 
with a short summary in chapter~IX.

\section{Kinks in $\mathbf{\phi^6}$ Models}

In $D=1+1$ dimensions the dynamics for the quantum field $\phi$ are governed solely 
by a field potential $U(\phi)$ that is added to the kinetic term
\begin{equation}
\mathcal{L}=\frac{1}{2}\partial_\mu \phi\partial^\mu \phi-U(\phi)\,.
\label{eq:lag1}
\end{equation}
For the $\phi^6$ model we scale all coordinates, fields and coupling constants
such that the potential contains only a single dimensionless parameter $a$
\begin{equation} 
U(\phi)=\frac{1}{2}\left(\phi^2+a^2\right)\left(\phi^2-1\right)^2\,.
\label{eq:pot1}
\end{equation}
\begin{figure}[t]
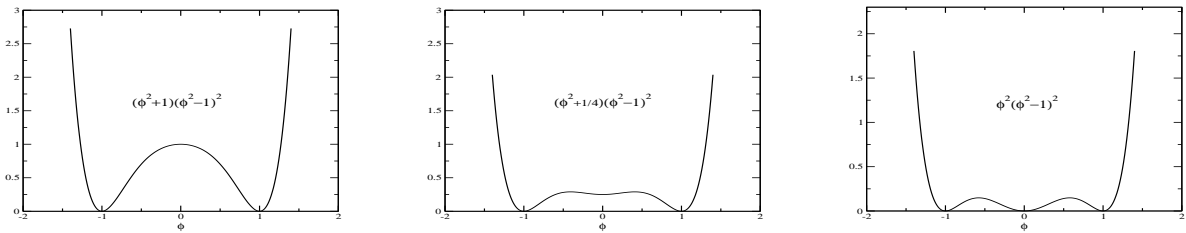

\centerline{\epsfig{file=p6a.eps,width=4.5cm,height=3cm}\hspace{1cm}
\epsfig{file=p6b.eps,width=4.5cm,height=3cm}\hspace{1cm}
\epsfig{file=p6c.eps,width=4.5cm,height=3cm}}
\caption{\label{fig:phi6pot}The field potential, eq.~(\ref{eq:pot1}) in the 
$\phi^6$ model for various values of the real parameter $a=1,\fract{1}{2},0$ from
left to right.}
\end{figure}

\begin{figure}[t]
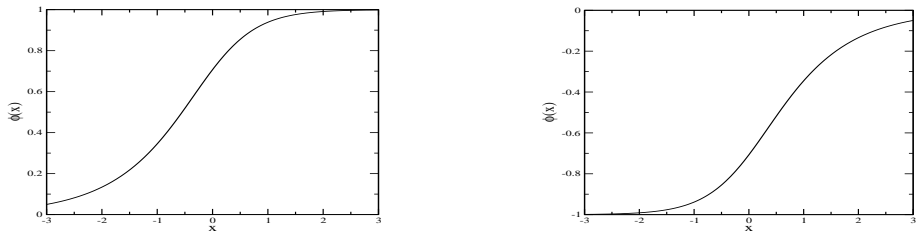

\centerline{
\epsfig{file=p6k1.eps,width=5cm,height=3.0cm}\hspace{2cm}
\epsfig{file=p6k2.eps,width=5cm,height=3.0cm}}
\caption{\label{fig:phi6sol}The two soliton solutions for $a=0$: Left panel:
eq~(\ref{eq:phi61}); right panel eq~(\ref{eq:phi62}).}
\end{figure}

\noindent
From figure~\ref{fig:phi6pot} we observe that there are three general cases. For 
$a^2>\fract{1}{2}$ two degenerate minima at $\phi=\pm1$ exist. For $0<a^2\le\fract{1}{2}$
an additional local minimum emerges at $\phi=0$. Finally, for $a=0$ the three minima
at $\phi=0$ and $\phi=\pm1$ are degenerate. Soliton solutions connect different vacua 
between negative and positive spatial infinity. For $a\ne0$ the vacua are at
$\phi=\pm1$ and the corresponding soliton solution is\cite{Lohe:1979mh}
\begin{equation}
\phi_K(x)=a\frac{X-1}
{\sqrt{4X+a^2\left(1+X\right)^2}}
\qquad \mbox{with} \qquad X={\rm e}^{2\sqrt{1+a^2}\,x}\,.
\label{eq:phik6a}
\end{equation}
Its classical energy is 
$E_{\rm cl}(a)=\frac{2-a^2}{4}\sqrt{1+a^2}+\frac{4a^2+a^4}{8}\,
{\rm ln}\frac{\sqrt{1+a^2}+1}{\sqrt{1+a^2}-1}$. The case $a=0$ is 
actually more interesting because two distinct soliton solutions do exist. 
The first one connects $\phi=0$ at $x\to-\infty$ to $\phi=1$ at $x\to\infty$,
\begin{equation}
\phi_{K_1}(x)=\frac{1}{\sqrt{1+{\rm e}^{-2x}}}\,,
\label{eq:phi61}
\end{equation}
while the second one interpolates between $\phi=-1$ and $\phi=0$,
\begin{equation}
\phi_{K_2}(x)=-\frac{1}{\sqrt{1+{\rm e}^{2x}}}\,.
\label{eq:phi62}
\end{equation}
These soliton configurations are shown in figure \ref{fig:phi6sol}.
In either case the classical mass is 
$E_{\rm cl}=\fract{1}{4}=\fract{1}{2}\lim_{a\to0}E_{\rm cl}(a)$. This 
relation for the classical energies reflects the fact that as $a\to0$ 
the solution $\phi_K(x)$ disintegrates into two widely separated structures 
one corresponding to $\phi_{K_1}(x)$ the other to $\phi_{K_2}(x)$.

The computation of the VPE requires the construction of scattering solutions for 
fluctuations about the soliton. In the harmonic approximation the fluctuations 
experience the potential 
\begin{equation}
V(x)=\frac{1}{2}\frac{\partial^2 U(\phi)}{\partial\phi^2}
\Big|_{\phi=\phi_{\rm sol}(x)}
\label{eq:fltpot1}
\end{equation}
generated by the soliton ($\phi_{\rm sol}=\phi_K$, $\phi_{K_1}$ or $\phi_{K_2}$).
\begin{figure}[t]
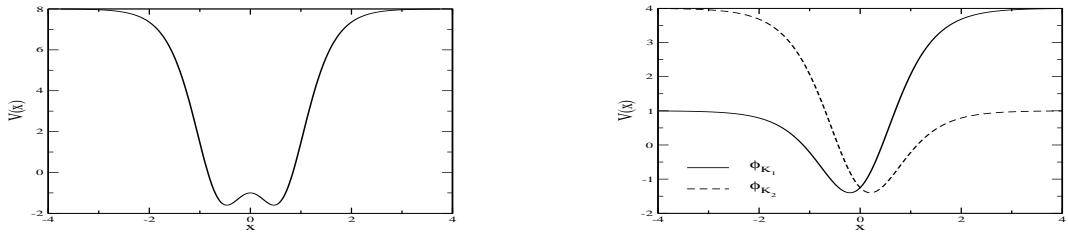

\centerline{
\epsfig{file=phi6a.eps,width=6cm,height=3cm}\hspace{2cm}
\epsfig{file=phi60.eps,width=6cm,height=3cm}}
\caption{\label{fig:fltpot}Scattering potentials for the 
quantum fluctuations in the $\phi^6$ model. Left panel: typical
example for $a\ne0$; right panel: the case $a=0$ with the two potentials
generated by $\phi_{K_1}$, full line and $\phi_{K_2}$, dashed line.}
\end{figure}
These three potentials are shown in figure \ref{fig:fltpot}. For
$a\ne0$ the potential is invariant under $x\leftrightarrow-x$. But
the particular case $a\equiv0$ is not reflection symmetric, though 
$x\leftrightarrow-x$ swaps the potentials generated by $\phi_{K_1}$ 
and $\phi_{K_2}$. The loss of this invariance disables the separation of 
the fluctuation modes into symmetric and anti--symmetric channels, which
is the one dimensional version of partial wave decomposition. Even more
strikingly, the different topological structures in the $a=0$ case 
cause $\lim_{x\to-\infty}V(x)\ne\lim_{x\to\infty}V(x)$, which implies 
different masses (dispersion relations) for the fluctuations at positive
and negative spatial infinity.

\section{Spectral Methods and Vacuum Polarization Energy}

The formula for the VPE, Eq.~(\ref{eq:master}) below, can be derived from first 
principles in quantum field theory by integrating the vacuum matrix element of the 
energy density operator\cite{Graham:2002xq}. It is, however, also illuminative to 
count the energy levels when summing the changes of the zero point energies. This 
sum is $\mathcal{O}(\hbar)$ and thus one loop order ($\hbar=1$ for the units used 
here). We call the single particle energies of fluctuations in the soliton type 
background $\omega_n$ while the $\omega_n^{(0)}$ are those for the trivial 
background. Then the VPE formally reads 
\begin{equation}
E_{\rm vac}=\frac{1}{2}\sum_n\left(\omega_n-\omega_n^{(0)}\right)
\Bigg|_{\rm ren.}
=\frac{1}{2}\sum_j \epsilon_j + 
\frac{1}{2} \int_0^\infty dk\, \omega_k\,\Delta\,\rho_{\rm ren.}(k)\,,
\label{eq:sum0}
\end{equation}
where the subscript indicates that renormalization is required to obtain a finite 
and meaningful result. On the right hand side we have separated the explicit bound 
state (sum of energies $\epsilon_j$) and continuum (integral over momentum $k$) 
contributions.  The latter involves $\Delta\,\rho_{\rm ren.}(k)$ which is the 
(renormalized) change of the level density induced by the soliton background. Let 
$L$ be a large distance away from the localized soliton background. For $x\sim L$ 
the stationary wave--function of the quantum fluctuation is a phase shifted plane 
wave $\psi(x)\sim{\rm sin}\left[kx+\delta(k)\right]$, where $\delta(k)$ is the phase 
shift (of a particular partial wave) that is obtained from scattering off the potential,
Eq.~(\ref{eq:fltpot1}). The continuum levels are counted from the boundary condition
$\psi(L)=0$ and subsequently taking the limit $L\to\infty$. The number $n(k)$ of levels 
with momentum less or equal to $k$ is then extracted from $kL+\delta(k)=n(k)\pi$. 
The corresponding number in the absence of the soliton is $n^{(0)}(k)=kL/\pi$, trivially. 
From these the change of the level density is computed via
\begin{equation}
\Delta\,\rho(k)=\lim_{L\to\infty}\frac{d}{dk}\left[n(k)-n^{(0)}(k)\right]
=\frac{1}{\pi}\frac{d\delta(k)}{dk}\,,
\label{eq:KFL}
\end{equation}
which is often referred to as the Krein--Friedel--Lloyd formula\cite{Faulkner:1977aa}.
Note that $\Delta\,\rho(k)$ is a finite quantity; but ultra--violet divergences appear
in the momentum integral in Eq.~(\ref{eq:sum0}) and originate from the large $k$ behavior
of the phase shift. This behavior is governed by the Born series
\begin{equation}
\delta(k)=\delta^{(1)}(k)+\delta^{(2)}(k)+\ldots
\label{eq:born}
\end{equation}
where the superscript reflects the power at which the potential, Eq.~(\ref{eq:fltpot1}) 
contributes. Though this series does not converge\footnote{For example, in three space
dimensions the series yields $\delta(0)\to0$ which contradicts Levinson's theorem.} for all 
$k$, it describes the large $k$ behavior well since
$\delta^{(N+1)}(k)/\delta^{(N)}(k)\propto 1/k^2$ when $k\to\infty$. Hence replacing 
\begin{equation}
\Delta\,\rho(k)\to\left[\Delta\,\rho(k)\right]_N=
\frac{1}{\pi}\frac{d}{dk}\left[\delta(k)-\delta^{(1)}(k)-\delta^{(2)}(k)-\ldots-
\delta^{(N)}(k)\right]
\label{eq:born1}
\end{equation}
produces a finite integral in Eq.~(\ref{eq:sum0}) when $N$ is taken sufficiently large.
We have to add back the subtractions that come with this replacement. Here the spectral 
methods take advantage of the fact that each term in the subtraction is uniquely related 
to a power of the background potential and that Feynman diagrams represent an alternative 
expansion scheme for the vacuum polarization energy
\begin{equation}
\mbox{\parbox[l]{2.2cm}{\vskip-1.9cm $E_{\rm FD}^{N}[V]=$}}
\epsfig{file=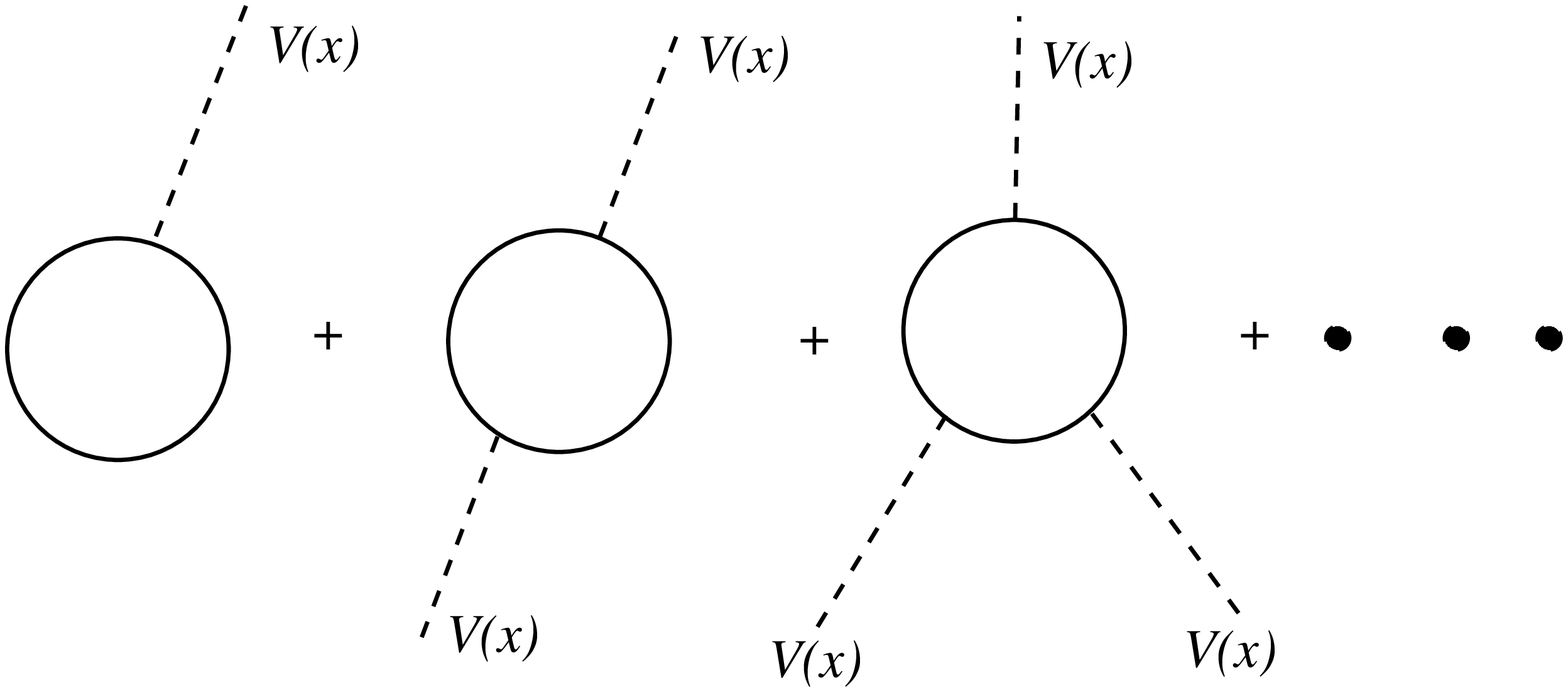,height=1.9cm,width=8cm}\,.
\label{eq:FDs}
\end{equation}
The full lines are the free propagators of the quantum fluctuations and the dashed lines 
denote insertions of the background potential, Eq.~(\ref{eq:fltpot1}), eventually after
Fourier transformation. These Feynman diagrams are regularized with standard techniques,
most commonly in dimensional regularization. They can thus be straightforwardly combined 
with the counterterm contribution, $E_{\rm CT}[V]$ with coefficients fully 
determined in the perturbative sector of the theory. This combination remains finite when 
the regulator is removed.  

The generalization to multiple channels is straightforward by finding an eventually 
momentum dependent diagonalization of the scattering matrix $S(k)$ and summing the 
so--obtained eigenphase shifts. This replaces\footnote{The proper Riemann sheet of the 
the logarithm is identified by constructing a smooth function that vanishes as $k\to\infty$.} 
$\delta(k)\,\longrightarrow\,(1/2\imu){\rm ln}{\rm det}\,S(k)$ and analogously for 
the Born expansions, Eqs.~(\ref{eq:born}) and~(\ref{eq:born1}). Since after
Born subtraction the integral converges, we integrate by parts to avoid numerical 
differentiation and to stress that the VPE is measured with respect to the translationally 
invariant vacuum.  We then find the renormalized VPE to be, with the sum over partial 
waves re--inserted,
\begin{equation}
E_{\rm vac}[V]=\sum_\ell D_\ell\left\{\frac{1}{2}\sum_j\left(\epsilon_{\ell j}-m\right)
- \int_0^\infty \frac{dk}{4\pi\imu} \frac{k}{\sqrt{k^2+m^2}}\,
\left[{\rm ln}\,{\rm det}\,S(k)\right]_{N}\right\}+E_{\rm FD}^{N}[V]+E_{\rm CT}[V]\,.
\label{eq:master}
\end{equation}
Here $D_\ell$ is the degree of degeneracy, {\it e.g.} $D_\ell=2\ell+1$ in three space
dimensions. The subscript $N$ refers to the subtraction of $N$ terms of the Born 
expansion, as {\it e.g.} in Eq.~(\ref{eq:born1}). We stress that, with $N$ taken sufficiently 
large, both the expression in curly brackets and the sum 
$E_{\rm FD}^{N}[V]+E_{\rm CT}[V]$ are individually ultra--violet finite and no 
cut--off parameter is needed\cite{Farhi:1998vx}.

\section{Scattering Data in One Space Dimension}

In this section we obtain the scattering matrix for general one dimensional problems
and develop an efficient method for its numerical evaluation. This will be at the center
of the novel approach to compute the VPE. 

We first review the standard approach that is applicable when $V(-x)=V(x)$, {\it e.g.} 
left panel of figure \ref{fig:fltpot}. Then the partial wave decomposition separates
symmetric $\psi_S(-x)=\psi_S(x)$ and anti--symmetric, $\psi_A(-x)=-\psi_A(x)$ channels.
The respective phase shifts can be straightforwardly obtained in a variant of the variable
phase approach\cite{Calegero:1967} by parameterizing 
$\psi(x)={\rm e}^{{\rm i}[kx+\beta(k,x)]}$ and imposing the obvious boundary conditions
$\psi^\prime_S(0)=0$ and $\psi_A(0)=0$. (The prime denotes the derivative with respect to
$x$.) The wave--equation turns into a non--linear differential equation for the phase function 
$\beta(k,x)$. When solved subject to 
$\lim_{x\to\infty}\beta(k,x)=0$ and $\lim_{x\to\infty}\beta^\prime(k,x)=0$
the scattering matrix given by\cite{Graham:2009zz}
\begin{equation}
\frac{1}{2{\rm i}}\,{\rm ln}\,{\rm det}\, S(k)=
-2{\sf Re}[\beta(k,0)]
-{\rm arctan}\frac{{\sf Im}[\beta^\prime(k,0)]}{k+{\sf Re}[\beta^\prime(k,0)]}\,.
\label{eq:sym1}
\end{equation}
Linearizing and iterating the differential equation for $\beta(k,x)$ yields the Born 
series, Eq.~(\ref{eq:born}). At this point it is advantageous to use the fact that 
scattering data can be continued to the upper half complex momentum plane\cite{Newton:1982qc}. 
That is, when writing $k=\imu t$, the Jost function, whose phase is the scattering phase shift 
when $k$ is real, is analytic for ${\sf Re}[t]\ge0$. Furthermore the Jost function has simple 
zeros at imaginary $k=\imu\kappa_j$ representing the bound states. Formulating the momentum 
integral from Eq.~(\ref{eq:master}) as a contour integral automatically collects the bound state 
contribution and we obtain a formula as simple as\cite{Graham:2002xq,Graham:2009zz}
\begin{equation}
E^{\rm (S)}_{\rm vac}=\int_{m}^\infty \frac{dt}{2\pi}\,
\frac{t}{\sqrt{t^2-m^2}}\,\left[
{\rm ln}\left\{g(t,0)\left(g(t,0)-\frac{1}{t}g^\prime(t,0)\right)\right\}
\right]_N +E_{\rm FD}^{N}[V]+E_{\rm CT}[V]
\label{eq:EvacJost}
\end{equation}
for the VPE.  Here $g(t,x)$ is the non--trivial factor of the Jost solution whose $x\to0$ 
properties determine the Jost function. The factor function solves the differential equation
\begin{equation}
g^{\prime\prime}(t,x)=2tg^\prime(t,x)+V(x)g(t,x)\,,
\label{eq:DEQJost}
\end{equation}
with the boundary conditions $g(t,\infty)=1$ and $g^\prime(t,\infty)=0$; iterating
$g(t,x)=1+g^{(1)}(t,x)+g^{(2)}(t,x)+\ldots$ produces the Born series.

In general, however, the potential $V(x)$ is not reflection invariant and no 
partial wave decomposition is applicable. Even more, there may exist different 
masses for the quantum fluctuations
\begin{equation}
m_L^2=\lim_{x\to-\infty}V(x)\qquad {\rm and}\qquad
m_R^2=\lim_{x\to\infty}V(x)
\label{eq:mass}
\end{equation}
as it is the case for the $\phi^6$ model with $a=0$, {\it cf.} right panel of 
figure \ref{fig:fltpot}. We adopt the convention that $m_L\le m_R$,
otherwise we simply swap $x\to-x$. Three different cases must be considered.
First, above threshold both momenta $k$ and $q=\sqrt{k^2+m_L^2-m_R^2}$ are real. To 
formulate the variable phase approach we introduce the matching point $x_m$ and 
parameterize
\begin{align}
x\le x_m:&\quad \psi(x)=A(x){\rm e}^{{\rm i}kx}\quad 
& A^{\prime\prime}(x)=-2{\rm i}kA^\prime(x)+V_p(x)A(x)\,\,\,\cr
x\ge x_m:&\quad \psi(x)=B(x){\rm e}^{{\rm i}qx}\quad 
& B^{\prime\prime}(x)=-2{\rm i}qB^\prime(x)+V_p(x)B(x)\,.
\label{eq:match1}
\end{align}
Observe that the {\it pseudo potential} 
\begin{equation}
V_p(x)=V(x)-m_L^2+(m_L^2-m_R^2)\Theta(x-x_m)
\label{eq:pseudoV}
\end{equation}
vanishes at positive and negative spatial infinity. The differential 
equations~(\ref{eq:match1}) are solved for the boundary conditions
conditions $A(-\infty)=B(\infty)=1$ and $A^\prime(-\infty)=B^\prime(\infty)=0$.
There are two linearly independent solutions $\psi_1$ and $\psi_2$ that define
the scattering matrix $S=(s_{ik})$ via the asymptotic behaviors
\begin{equation}
\psi_1(x)\sim \begin{cases}
{\rm e}^{{\rm i}kx}+s_{12}(k){\rm e}^{-{\rm i}kx}\quad &{\rm as}\quad x\to-\infty\cr
s_{11}(k){\rm e}^{{\rm i}qx}\quad &{\rm as}\quad x\to\infty
\end{cases}
\hspace{0.7cm}{\rm and}\hspace{0.7cm}
\psi_2(x)\sim \begin{cases}
s_{22}(k){\rm e}^{-{\rm i}kx}\quad &{\rm as}\quad x\to-\infty\cr
{\rm e}^{-{\rm i}qx}+s_{21}(k){\rm e}^{{\rm i}qx}\quad &{\rm as}\quad x\to\infty\,.
\end{cases}
\hspace{0.7cm}
\label{eq:defS}
\end{equation}
By equating the solutions and their derivatives at $x_m$ the scattering matrix is 
obtained from the factor functions as
\begin{align}
S(k)=&\begin{pmatrix}
{\rm e}^{-\imu qx_m} & 0 \cr 
0 & {\rm e}^{\imu kx_m}
\end{pmatrix}
\begin{pmatrix}
B & -A^\ast \cr
iqB+B^\prime & ikA^\ast-A^{\prime\ast}
\end{pmatrix}^{-1}\cr&\hspace{1cm}\times
\begin{pmatrix}
A & -B^\ast \cr
ikA+A^\prime & iqB^\ast-B^{\prime\ast}
\end{pmatrix}
\begin{pmatrix}
{\rm e}^{\imu kx_m} & 0 \cr 
0 & {\rm e}^{-\imu qx_m}
\end{pmatrix}
\hspace{2cm} {\rm for}\quad k\ge\sqrt{m_R^2-m_L^2}\,,
\label{eq:Smat1}
\end{align}
where $A=A(x_m)$, etc.. The second case refers to $k\le\sqrt{m_R^2-m_L^2}$ still being 
real but $q=\imu\kappa$ becoming imaginary with $\kappa=\sqrt{m_R^2-m_L^2-k^2}$. The parameterization
of the wave function for $x>x_m$ changes to $\psi(x)=B(x){\rm e}^{-\kappa x}$ yielding the
differential equation $B^{\prime\prime}(x)=\kappa B^\prime(x)+V_p(x)B(x)$. The scattering matrix
then is a single unitary number
\begin{equation}
S(k)=-\,\frac{A\left(B^\prime/B-\kappa-ik\right)-A^\prime}
{A^\ast\left(B^\prime/B-\kappa+ik\right)-A^{\prime\ast}}\,
{\rm e}^{2\imu kx_m}
\hspace{2cm} {\rm for}\quad 0\le k\le\sqrt{m_R^2-m_L^2}\,.
\label{eq:Smat2}
\end{equation}
It is worth noting that $V_p\equiv0$ corresponds to the step function potential. In that case 
the above formalism obviously yields $A\equiv B\equiv1$ and reproduces the textbook result
\begin{equation}
\delta_{\rm step}(k)=
\begin{cases}
(k-q)x_m\,,\quad & {\rm for}\quad k\ge\sqrt{m_R^2-m_L^2}\cr 
kx_m-{\arctan}\left(\frac{\sqrt{m_R^2-m_L^2-k^2}}{k}\right)\,,
\quad & {\rm for}\quad k\le\sqrt{m_R^2-m_L^2}\,.
\end{cases}
\label{eq:step1}
\end{equation}
In the third regime also $k$ becomes imaginary and we need to identify the bound
states energies $\epsilon\le m_L$ that enter Eq.~(\ref{eq:master}). We define real
variables $\lambda=\sqrt{m_L^2-\epsilon^2}$ and $\kappa(\lambda)
=\sqrt{m_R^2-m_L^2+\lambda^2}$ and solve the wave equation subject to the initial conditions
\begin{equation}
\psi_L(x_{\rm min})=1\,,\qquad
\psi^\prime_L(x_{\rm min})=\lambda
\qquad {\rm and}\qquad
\psi_R(x_{\rm max})=1\,,\qquad
\psi^\prime_R(x_{\rm max})=-\kappa(\lambda)\,,
\label{eq:bound1}
\end{equation}
where $x_{\rm min}$ and $x_{\rm max}$ represent negative and positive spatial infinity,
respectively. Continuity of the wave function requires the Wronskian determinant
\begin{equation}
\psi_L(x_m)\psi^\prime_R(x_m)-\psi_R(x_m)\psi^\prime_L(x_m)\stackrel{!}{=}0\,,
\label{eq:bound2}
\end{equation}
to vanish. This occurs only for discrete values $\lambda_j$ that in turn determine 
the bound state energies\footnote{The bosonic dispersion relation does not exclude 
imaginary energies that would hamper the definition of the quantum theory. This case 
does not occur here.} $\epsilon_j=\sqrt{m_L^2-\lambda_j^2}$.

\section{One Loop Renormalization in One Space Dimension}

To complete the computation of the VPE we need to substantiate the renormalization
procedure. We commence by identifying the ultra--violet singularities. This is simple
in $D=1+1$ dimensions at one loop order as only the first diagram on the right hand side 
of Eq.~(\ref{eq:FDs}) is divergent. Furthermore, this diagram is local in the sense
that $E_{\rm FD}^{(1)}\propto \frac{1}{\epsilon}\int dx\,\left[V(x)-m_L^2\right]$,
where $\epsilon$ is the regulator ({\it e.g.} from dimensional regularization). Hence 
a counterterm can be constructed that not only removes the singularity but the
diagram in total. This is the so--called {\it no tadpole} condition and implies
\begin{equation}
E_{\rm FD}^{(1)}+E_{\rm CT}^{(1)}=0\,.
\label{eq:notad}
\end{equation}
In the next step we must identify the corresponding Born term in Eq.~(\ref{eq:born}).
To this end it is important to note that the counterterm is a functional of the 
full field $\phi(x)$ that induces the background potential, Eq.~(\ref{eq:fltpot1}).
Hence we must find the Born approximation for $V(x)-m_L^2$ rather than the one for the 
pseudo--potential $V_P(x)$, Eq.~(\ref{eq:pseudoV}). The standard formulation of the Born 
approximation as an integral over the potential is, unfortunately, not applicable to 
$V(x)-m_L^2$ since it does not vanish at positive spatial infinity. However, we note that 
$V(x)-m_L^2=V_P(x)+(m_L^2-m_R^2)\Theta(x-x_m)=V_p(x)+V_{\rm step}(x)$ and that, by
definition, the first order correction is linear in the background, and thus additive. 
We may therefore write
\begin{equation}
\delta^{(1)}(k)=\delta^{(1)}_P(k)+\delta^{(1)}_{\rm step}(k)
=\frac{-1}{2k}\int_{-\infty}^\infty dx\,
V_p(x)\Big|_{x_m}+\frac{x_m}{2k}\left(m_L^2-m_R^2\right)
=\frac{-1}{2k}\int_{-\infty}^\infty dx\, V_p(x)\Big|_{0}\,.
\label{eq:born2}
\end{equation}
The Born approximation for the step function potential has been obtained from 
the large $k$ expansion of $\delta_{\rm step}(k)$ in Eq.~(\ref{eq:step1}). The 
subscripts in Eq.~(\ref{eq:born2}) recall that the definition of the 
pseudo--potential, Eq.~(\ref{eq:pseudoV}) induces an implicit dependence on the 
(artificial) matching point $x_m$. Notably, this dependence disappears from the
final result! This is the first step towards establishing the matching point
independence of the VPE.

The integrals in $E_{\rm FD}^{(1)}$ and $E_{\rm CT}^{(1)}$ require further 
regularization when $m_L\ne m_R$. In that case no further {\it finite 
renormalization} beyond the no tadpole condition is realizable.

\section{Comparison with Known Results}

Before presenting detailed numerical results for VPEs, we note that all 
simulations were verified to produce $S^\dagger S=\ID$ after attaching pertinent 
flux factors to the scattering matrix, Eq.~(\ref{eq:defS}). These flux factors
are not relevant for the VPE as they multiply to unity under the determinant 
in Eq.~(\ref{eq:master}). In addition the numerically obtained phase shifts, 
{\it i.e.} $(1/2\imu){\rm ln}{\rm det}\,S$, have been monitored to not vary with 
$x_m$. Since this is also the case for the bound energies, the VPE is verified to
be independent of the unrestricted choice for the matching point.

The VPE calculation based on Eq.~(\ref{eq:master}) has been applied to the 
$\phi^4$ kink and sine--Gordon soliton models that are defined via the potentials
\begin{equation}
U_K(\phi)=\fract{1}{2}\left(\phi^2-1\right)^2
\qquad {\rm and}\qquad
U_{\rm SG}(\phi)=4\left({\rm cos}(\phi)-1\right)\,,
\label{eq:known1}
\end{equation}
respectively. The soliton solutions $\phi_K={\rm tanh}(x-x_0)$ and
$\phi_{\rm SG}(x)=4{\rm arctan}\left({\rm e}^{-2(x-x_0)}\right)$ induce the 
scattering potentials
\begin{equation}
V_K(x)-m^2=6\left[{\rm tanh}^2(x-x_0)-1\right]
\qquad {\rm and}\qquad
V_{\rm SG}(x)-m^2=8\left[{\rm tanh}^2[2(x-x_0)]-1\right]\,.
\label{eq:known2}
\end{equation}
In both cases we have identical dispersion relations at positive and negative spatial 
infinity: $m=m_L=m_R=2$ for the dimensionless units introduced above. The simulation 
based on Eq.~(\ref{eq:master}) reproduces the established results 
$E_{\rm vac}^{(K)}=\fract{\sqrt{2}}{4}-\fract{3}{\pi}$ and 
$E_{\rm vac}^{({\rm SG})}=-\fract{2}{\pi}$\cite{Ra82}. These solitons break translational
invariance spontaneously and thus produce zero mode bound states in the fluctuation spectrum. 
In addition the $\phi^4$ kink possesses a bound state with energy $\sqrt{3}$\cite{Ra82}. 
All bound states are easily observed using Eq.~(\ref{eq:bound2}). The potentials in 
Eq.~(\ref{eq:known2}) are reflection symmetric about the soliton center $x_0$ and the method 
of Eq.~(\ref{eq:EvacJost}) can straightforwardly applied\cite{Graham:2009zz}. However, this 
method singles out $x_0$ (typically set to $x_0=0$) to determine the boundary condition in the 
differential equation and therefore cannot be used to establish translational invariance of 
the VPE.  On the contrary, the boundary conditions for Eq.~(\ref{eq:match1}) are not at all 
sensitive to $x_0$ and we have applied the present method to compute the VPE for various 
choices of $x_0$, all yielding the same numerical result.

The next step is to compute the VPE for asymmetric background potentials that have $m=m_L=m_R$. 
For the lack of a soliton model that produces such a potential we merely consider a two 
parameter set of functions
\begin{equation}
V_p(x)\,\longrightarrow\,V_{R,\sigma}(x)=Ax{\rm e}^{-x^2/\sigma^2}
\label{eq:asym1}
\end{equation}
for the pseudo potential in Eq.~(\ref{eq:match1}). Although Eq.~(\ref{eq:EvacJost}) is not 
directly applicable, it is possible to relate $V_{R,\sigma}(x)$ to the symmetric potential
\begin{equation}
V_R(x)=A\left[(x+R){\rm e}^{-\frac{(x+R)^2}{\sigma^2}}
-(x-R){\rm e}^{-\frac{(x-R)^2}{\sigma^2}}\right]=V_R(-x)
\label{eq:asym2}
\end{equation}
and apply Eq.~(\ref{eq:EvacJost}). In the limit $R\to\infty$ interference effects between the 
two structures around $x=\pm R$ disappear resulting in twice the VPE of Eq.~(\ref{eq:asym1}). 
The numerical comparison is listed in table \ref{tab:asym}.
\begin{table}[t]
\centerline{
\begin{tabular}{c|cccccc|c}
$R$ & 1.0 & 1.5 & 2.0 & 2.5 & 3.0 & 3.5 & present \cr
\hline
$A=2.5\,,\,\sigma=1.0$ &
-0.0369 & -0.0324 & -0.0298 & -0.0294 & -0.0293 & -0.0292 & -0.0293 \cr
\hline\hline
$R$ & 4.0 & 5.0 & 6.0 & 7.0 & 8.0 & 9.0 & present \cr
\hline
$A=0.2\,,\,\sigma=4.0$ &
-0.0208 & -0.0188 & -0.0170 & -0.0161 & -0.0158 & -0.0157 & -0.0157
\end{tabular}}
\caption{\label{tab:asym}The $R$ dependent data are half the VPE for the 
symmetrized potential, Eq.~(\ref{eq:asym2}) computed from Eq.~(\ref{eq:EvacJost}).
The data in the column {\it present} list the results obtained from 
Eq.~(\ref{eq:master}) for the original potential, Eq.~(\ref{eq:asym1}).}
\end{table}
Indeed the two approaches produce identical results as $R\to\infty$. 
The symmetrized version converges only slowly for wide potentials
(large $\sigma$) causing obstacles for the numerical simulation that
do not at all occur in the present approach.

\section{Vacuum Polarization Energies in the $\mathbf{\phi^6}$ Model}

We first discuss the VPE for the $a\ne0$ case. A typical background potential
is shown in the left panel of figure \ref{fig:phi6pot}. Obviously it is reflection
invariant and thus the method based on Eq.~(\ref{eq:EvacJost}) is applicable. In 
table \ref{tab:phi6a} we also compare our results to those from the heat kernel 
expansion of Ref.\cite{AlonsoIzquierdo:2011dy} since, to our knowledge, it is 
the only approach that has also been applied to the asymmetric $a=0$ case in 
Ref.\cite{AlonsoIzquierdo:2002eb}.
\begin{table}[t]
\centerline{
\begin{tabular}{c|ccccccc}
$a$ & 0.001 & 0.01 & 0.05 & 0.1 & 0.2 & 1.0 & 1.5 \cr
\hline
heat kernel, Ref.\cite{AlonsoIzquierdo:2011dy}~~ 
& -1.953  & -1.666 & -1.447 & -1.349 
& -1.239 & -1.101 & -1.293\cr
parity sep., Eq.~(\ref{eq:EvacJost})~~ 
& -2.145 & -1.840 & -1.595 & -1.461 &
-1.298 & -1.100 & -1.295 \cr
present, Eq.(\ref{eq:master}) 
& -2.146 & -1.841 & -1.596 & -1.462 
& -1.297 & -1.102 & -1.297
\end{tabular}}
\caption{\label{tab:phi6a}Different methods to compute the VPE of 
the $\phi^6$ soliton for $a\ne0$.}
\end{table}
Not surprisingly, the two methods based on scattering data agree within
numerical precision for all values of $a$. The heat kernel results also 
agree for moderate and large~$a$; but for small values deviations of the order 
of 10\% are observed. The heat kernel method relies on truncating the 
expansion of the exact heat kernel about the heat kernel in the absence 
of a soliton. Although in Ref.\cite{AlonsoIzquierdo:2011dy} the expansion has 
been carried out to eleventh(!) order, leaving behind a very cumbersome 
calculation, this does not seem to provide sufficient accuracy for small $a$.

\begin{figure}[t]
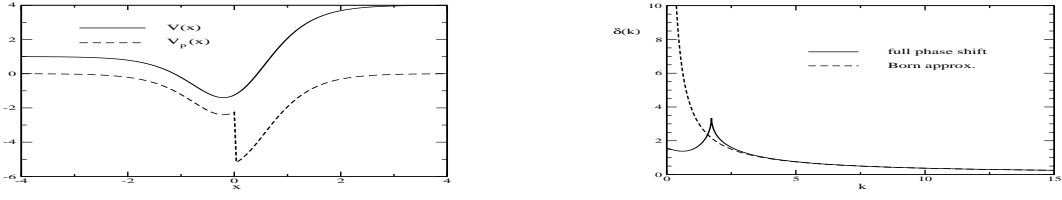

\centerline{
\epsfig{file=pot.eps,width=6cm,height=2.5cm}\hspace{2cm}
\epsfig{file=delta.eps,width=6cm,height=2.5cm}}
\caption{\label{fig:phi6}Left panel: potential ($V$) and pseudo potential ($V_p$)
for fluctuations about a $\phi^6$ soliton with $a=0$. The pseudo potential
is shown for $x_m=0$. Right panel: resulting phase shift, {\it i.e.} 
$(1/2\imu) {\rm ln}{\rm det}\, S$, full line and its Born approximation,
dashed line.}
\end{figure}

We are now in the position to discuss the VPE for $a=0$ associated with the
soliton $\phi_{K_1}(x)$ from Eq.~(\ref{eq:phi61}). The potentials for the
fluctuations and the resulting scattering data are shown in 
figure \ref{fig:phi6}. By construction, the pseudo potential jumps at $x_m=0$.
However, neither the phase shift nor the bound state energy (the zero mode
is the sole bound state) depends on $x_m$.
As expected, the phase shift has a threshold cusp at
$\sqrt{m_R^2-m_L^2}=\sqrt{3}$ and approaches $\frac{\pi}{2}$ at zero momentum. 
This is consistent with Levinson's theorem in one space dimension\cite{Barton:1984py}
and the fact that there is only a single bound state. In total we find a significant
cancellation between bound state and continuum contributions
\begin{equation}
E_{\rm vac}=-0.5+0.4531=-0.0469\,.
\label{eq:main}
\end{equation}
The result\footnote{The factor $\sqrt{2}$ is added to adjust the datum 
from Ref.\cite{AlonsoIzquierdo:2002eb} to the present scale.}  
$-0.1264\sqrt{2}=-0.1788$ of Ref.\cite{AlonsoIzquierdo:2002eb} 
was estimated relative to $V_\alpha(x)=\frac{3}{2}\left[1+{\rm tanh}(\alpha x)\right]$
for $\alpha=1$. Our results for various values of $\alpha$ are listed in
table \ref{tab:tanh}. These results are consistent with $V_\alpha(x)$ turning into
a step function for large $\alpha$. For the particular value $\alpha=1$ our relative
VPE thus is $\Delta E_{\rm vac}=-0.0469-0.1660=-0.2129$. In
view of the results shown in table \ref{tab:phi6a}, especially for small $a$,
these data match within the validity of the approximations applied in the
heat kernel calculation.
\begin{table}[t]
\centerline{
\begin{tabular}{c|ccccc|c}
$\alpha$ & 1.0 & 2.0 & 5.0 & 10.0 & 30.0 & step\cr
\hline
$E_{\rm vac}$& 0.1660 & 0.1478 & 0.1385 & 0.1363 & 0.1355 & 0.1355
\end{tabular}}
\caption{\label{tab:tanh}VPE for background potential $V_\alpha(x)$ defined 
in the main text. The entry {\it step} gives the VPE for the step function
potential $V(x)=3\Theta(x)$ using Eq.~(\ref{eq:step1}) and its Born approximation
from Eq.~(\ref{eq:born2}) for $x_m=0$.}
\end{table}

\section{Translational Variance}

So far we have computed the VPE for the $\phi^6$ model soliton centered at 
$x_0=0$. We have already mentioned that there is translational invariance 
for the VPE of the kink and sine--Gordon solitons. It is also numerically 
verified for the asymmetric background, Eq.~(\ref{eq:asym1}). In those cases
the two vacua at $x\to\pm\infty$ are equivalent and $q=k$ in Eq.~(\ref{eq:defS}).
When shifting $x\to x+x_0$, the transmission coefficients ($s_{11}$ and $s_{22}$) 
remain unchanged relative to the amplitude of the in--coming wave while the 
reflection coefficients ($s_{12}$ and $s_{21}$) acquire opposite phases. Consequently, 
${\rm det}\,S$ is invariant. For unequal momenta this invariance forfeits and the 
VPE depends on $x_0$. This is reflected by the results in 
table \ref{tab:shift} in which we present the VPE for 
$V_\alpha(x)=\frac{3}{2}\left[1+{\rm tanh}(\alpha (x+x_0))\right]$ and
the $\phi^6$ model soliton $1/\sqrt{1+{\rm e}^{-2(x+x_0)}}$.
\begin{table}[b]
\centerline{
\begin{tabular}{c|ccccc}
&\multicolumn{5}{c}{$E_{\rm vac}$}\cr
\hline
$x_0$& -2 & -1 & 0 & 1 & 2\cr
\hline
$\alpha=5$ & 0.341 & 0.240 & 0.139 & 0.037 & -0.064\cr
$\alpha=2$ & 0.351 & 0.250 & 0.148 & 0.046 & -0.057\cr
$\alpha=1$ & 0.369 & 0.267 & 0.166 & 0.064 & -0.038\cr
$\phi^6$ & 0.154 & 0.053 & -0.047 & -0.148 & -0.249\cr
$\Delta E_{\rm vac}$ & -0.215 & -0.214 & -0.213 & -0.212 & -0.211
\end{tabular}}
\caption{\label{tab:shift}The VPE as function of the position of the center
of the potential for $V_\alpha$ and the $\phi^6$ model soliton.
$\Delta E_{\rm vac}$ is the difference between the VPEs of 
the latter and $V_1$.}  
\end{table}
Obviously there is a linear dependence of the VPE on $x_0$ with the slope 
insensitive to specific structure of the potential. This insensitivity is 
consistent with the above remark on the difference between the two momenta.
Increasing $x_0$ shifts the vacuum with the bigger mass towards negative
infinity thereby removing states from the spectrum and hence decreasing
the VPE.

The effect is immediately linked to varying the width of a symmetric barrier
potential with height $m_R^2-m_L^2=3$:
\begin{equation}
V^{(x_0)}_{\rm SB}(x)=3\Theta\left(\frac{x_0}{2}-|x|\right)\,.
\label{eq:symbarr}
\end{equation}
For this potential the Jost solution, Eq.~(\ref{eq:DEQJost}) can be obtained
analytically\cite{Weigel:2016zbs} and the VPE has the limit
\begin{equation}
\lim_{x_0\to\infty}\frac{E_{\rm vac}[V^{(x_0)}_{\rm SB}]}{x_0}\approx-0.102\,,
\label{eq:barrlim}
\end{equation}
which again reveals the background independent slope observed above. 

\begin{figure}[t]
\centerline{\epsfig{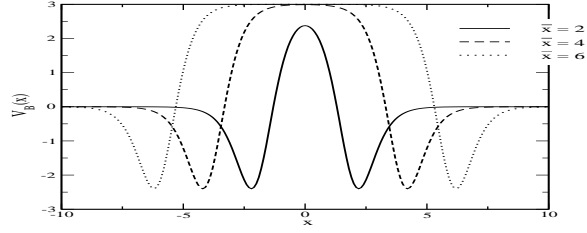}}
\caption{\label{fig:plotbg}Background potential for the kink--antikink pair, 
Eq.~(\ref{eq:pair}) for different separations.}
\end{figure}

Having quantitatively determined the translation variance of the VPE, it is 
tempting to subtract $E_{\rm vac}\left[V^{(x_0)}_{\rm SB}\right]$. Unfortunately 
this is not unique because $x_0$ is not the unambiguous center of the soliton. 
For example, employing the classical energy density $\epsilon(x)$ to define 
the position of the soliton $1/\sqrt{1+{\rm e}^{-2(x-\overline{x})}}$, that is 
formally centered at $\overline{x}$, as an expectation value leads to
\begin{equation}
x_s=\frac{\int dx x \epsilon(x)}{\int dx \epsilon(x)}=\overline{x}+\fract{1}{2}\,.
\end{equation}
This changes the VPE by approximately $0.050$. This ambiguity also hampers the
evaluation of the VPE as half that of a widely separated kink--antikink pair 
\begin{equation}
\phi_{K\overline{K}}(x)=\left[1+{\rm e}^{2(x-\overline{x})}\right]^{-1/2}
+\left[1+{\rm e}^{-2(x+\overline{x})}\right]^{-1/2}-1
\label{eq:pair}
\end{equation}
similarly to the approach for Eq.~(\ref{eq:asym2}). The corresponding 
background potential $V_B$ is shown in figure \ref{fig:plotbg}.
For computing the VPE, the large contribution from the constant but non--zero 
potential in the regime $|x|\lesssim \overline{x}$ should be eliminated. The above
considerations lead to
\begin{align}
&\fract{1}{2}\lim_{\bar{x}\to\infty}\left\{E_{\rm vac}[V_B]
-2E_{\rm vac}[V^{(2\overline{x})}_{\rm SB}]\right\}=-0.170
\quad{\rm and}\quad
\fract{1}{2}\lim_{\bar{x}\to\infty}\left\{E_{\rm vac}[V_B]
-2E_{\rm vac}[V^{(2x_s)}_{\rm SB}]\right\}=-0.120\,.
\end{align}
When the VPE from $V^{(2(\overline{x}+1.2)}_{\rm SB}$ is subtracted, the main 
result, Eq.~(\ref{eq:main}), is matched. Eventually this can be used to define 
the center of the soliton.

Now we also understand why the VPE for $a\ne0$ diverges as $a\to0$, {\it cf.} 
table \ref{tab:phi6a}. In that limit kink and antikink structures separate and 
the ''vacuum'' in between produces an ever increasing contribution (in magnitude).

Finally, we discuss the link between the translational variance and the 
Krein--Friedel--Lloyd formula, Eq.~(\ref{eq:KFL}). We have already reported
the VPE for the step function potential when $x_m=0$. We can also consider 
$x_m\to\infty$: 
\begin{align}
\frac{E_{\rm vac}[V^{(x_m)}_{\rm step}]}{|x_m|}\,\to\,&-{\rm sign}(x_m)\,
\left[\int_0^{\sqrt{3}}\frac{dk}{4\pi}\,\frac{2k^2-3}{\sqrt{k^2+1}}
+\int_{\sqrt{3}}^\infty\frac{dk}{4\pi}\,
\frac{2k^2-2k\sqrt{k^2-3}-3}{\sqrt{k^2+1}}\right]
\approx0.101\,{\rm sign}(x_m)\,,
\label{eq:xminf}
\end{align}
reproducing the linear dependence on the position from above. Formally, {\it i.e.}
without Born subtraction, the integral, Eq.~(\ref{eq:xminf}) is dominated by
\begin{align}
\int \frac{dk}{2\pi}\,\frac{k}{\sqrt{k^2+1}}\left[k-\sqrt{k^2-3}\right]
\sim \int \frac{dk}{2\pi}\, \sqrt{k^2+1} \frac{d}{dk}\left[\sqrt{k^2-3}-k\right]
=\int \frac{dk}{2\pi}\, \sqrt{k^2+1}\, \frac{d}{dk}\left[q-k\right]\,.
\end{align}
Essentially this is that part of the level density that originates from 
the different dispersion relations at positive and negative spatial infinity.

\section{Conclusion}

We have advanced the spectral methods for computing vacuum polarization 
energies (VPE) to also apply for static localized background configurations in 
one space dimension that do not permit a parity decomposition for the quantum 
fluctuations. The essential progress is the generalization of the variable 
phase approach to such configurations. Being developed from spectral methods, 
it adopts their amenities, as for {\it e.g.} an effective procedure to implement 
standard renormalization conditions. A glimpse at the bulky formulas for the 
heat kernel expansion (alternative method to the problem)  in
Refs.\cite{AlonsoIzquierdo:2002eb,AlonsoIzquierdo:2011dy,AlonsoIzquierdo:2012tw}
immediately reveals the simplicity and effectiveness of the present approach. The
latter merely requires to numerical integrate ordinary differential equations and 
extract the scattering matrix thereof, {\it cf.} Eqs.~(\ref{eq:match1}) 
and~(\ref{eq:Smat1}). Heat kernel methods are typically combined with $\zeta$--function
regularization. Then the connection to standard renormalization conditions is
not as transparent as for the spectral methods, though that is problematic only
when non--local Feynman diagrams require renormalization, {\it i.e.} in larger
than $D=1+1$ dimensions or when fermion loops are involved.

We have verified the novel method by means of well established results, as, {\it e.g.}
the $\phi^4$ kink and sine--Gordon solitons. For these models the approach directly
ascertains translational invariance of the VPE. Yet, the main focus was on the VPE for
solitons in $\phi^6$ models because its soliton(s) may connect in--equivalent vacua
leading to background potentials that are not invariant under spatial reflection. This 
model is not strictly renormalizable. Nevertheless at one loop order a well defined result 
can be obtained from the no--tadpole renormalization condition albeit no further finite 
renormalization is realizable because the different vacua yield additional infinities
when integrating the counterterm. The different vacua also lead to different dispersion 
relations for the quantum fluctuations and thereby induce translational variance for a 
theory that is formulated by an invariant action. We argue that this variance is universal, 
as it is not linked to the particular structure of the background and can be related to 
the change in the level density that is basic to the Krein--Friedel--Lloyd formula, 
Eq.~(\ref{eq:KFL}).

Besides attempting a deeper understanding of the variance by tracing it from the 
energy momentum tensor, future studies will apply the novel method to solitons of 
the $\phi^8$ model. Its elaborated structure not only induces potentials that are 
reflection asymmetric, but also leads to a set of topological indexes\cite{Gani:2015cda} 
that are related to different particle numbers. Then the novel method will
progress the understanding of quantum corrections to binding energies of compound 
objects in the soliton picture. Furthermore the present results can be joined with 
the interface formalism\cite{Graham:2001dy}, that augments additional coordinates 
along which the background is homogeneous, to explore the energy (densities) of
domain wall configurations\cite{Parnachev:2000fz}.

\section*{Acknowledgments}
This work was presented at the $5^{\rm th}$ {\it Winter Workshop on 
Non--Perturbative Quantum Field Theory}, Sophia-Antipolis (France), March 2017.
The author is grateful to the organizers for providing this worthwhile workshop. 

The author declares that there is no conflict of interest regarding the 
publication of this paper. This work is supported in parts by the NRF 
under grant~109497.

\end{document}